\documentclass{elsart}
\usepackage{graphicx}
\usepackage{epsfig}
\usepackage{amssymb}

\renewcommand{\d}{{\mathrm d}}
\renewcommand{\bar}[1]{\overline{#1}}

\providecommand{\Journal}[4] {#1 {\bf #2} (#4) #3}
\providecommand{\NIMA}{Nucl. Instr. Meth. A } %
\providecommand{\NPB}{Nucl. Phys. B } %
\providecommand{\PL}{Phys. Lett. } %
\providecommand{\PLB}{Phys. Lett. B } %
\providecommand{\PRL}{Phys. Rev. Lett. } %
\providecommand{\PRC}{Phys. Rev. C } %
\providecommand{\PRD}{Phys. Rev. D } %

\journal{Physics Letters B}

\begin{document}
\begin{frontmatter}

\begin{flushright}
Preprint USM-TH-140, published in Phys.Lett. B574 (2003) 35.
\end{flushright}

\title{Ratio of $\bar{\Lambda}/\Lambda$ in Semi-inclusive
Electroproduction}

\author[pku]{Bo-Qiang Ma},
\ead{mabq@phy.pku.edu.cn}
\author[utfsm]{Ivan Schmidt},
\ead{ischmidt@fis.utfsm.cl}
\author[utfsm,UR,NNU]{Jian-Jun Yang}
\ead{jjyang@fis.utfsm.cl}
\address[pku]{Department of Physics, Peking University, Beijing 100871, China}
\address[utfsm]{Departamento de F\'\i sica, Universidad
T\'ecnica Federico Santa Mar\'\i a, Casilla 110-V, Valpara\'\i so,
Chile}
\address[UR]{Institut f\"ur Theoretische Physik,
Universit\"at Regensburg, D-93040 Regensburg, Germany}
\address[NNU]{Department of Physics, Nanjing Normal
University, Nanjing 210097, China}



\begin{abstract}
It is shown that the $\bar{\Lambda}/\Lambda$ cross section ratio
in semi-inclusive electroproduction of $\Lambda$ and
$\bar{\Lambda}$ hyperons in deep inelastic scattering of charged
lepton on a nucleon target, can provide useful information on the
quark to $\Lambda$ fragmentation functions. This ratio is
calculated explicitly in a quark-diquark model, a pQCD based
analysis, and an SU(3) symmetry model, with three different
options for the contribution from the unfavored fragmentation
functions. The $x$-dependence of this ratio is sensitive to the
ratio of unfavored fragmentation functions over favored
fragmentation functions,
$D_{\bar{u}}^{\Lambda}(z)/D_{u}^{\Lambda}(z)$, whereas the
$z$-dependence is sensitive to the flavor structure of the
fragmentation functions, i.e., the ratio
$D_u^{\Lambda}(z)/D_s^{\Lambda}(z)$. Future measurements by the
HERMES Collaboration at DESY can discriminate between various
cases.
\end{abstract}


\end{frontmatter}



The structure of the $\Lambda$ hyperon is one of the main problems
of hadron physics, and is under active investigation both
theoretically \cite{MSSY01} and experimentally
\cite{ALEPH,OPAL,HERMES,E665,NOMAD,E769}. Due to its short life
time and neutral charge, the $\Lambda$ cannot be used as target
nor as beam, therefore it is difficult to measure the quark
distributions of the $\Lambda$ directly. The most effective means
to investigate the structure of the $\Lambda$ is through the quark
to $\Lambda$ fragmentation in various processes
\cite{MSY00PLB,MSY00Z}. However, there is still no available
experimental measurement about the relations between different
flavor-dependent quark to $\Lambda$ fragmentation functions, such
as the relation between the favored and unfavored fragmentation
functions $D_u^{\Lambda}(z)$, $D_s^{\Lambda}(z)$ and
$D_{\bar{u}}^{\Lambda}(z)$, $D_{\bar{s}}^{\Lambda}(z)$, and the
relation between the flavor structure of the favored fragmentation
functions $D_u^{\Lambda}(z)$ and $D_s^{\Lambda}(z)$.

The purpose of this letter is to show that the
$\bar{\Lambda}/\Lambda$ ratio of semi-inclusive electroproduction
cross sections in deep inelastic scattering of charged leptons on
a nucleon target is a sensitive physical quantity which can
provide useful information about the relations between different
quark to $\Lambda$ fragmentation functions. We will calculate the
$\bar{\Lambda}/\Lambda$ cross section ratio in three different
models: a quark-diquark model \cite{MSY00PLB,MSSY00}, a
perturbative QCD (pQCD) based analysis \cite{MSY00PLB,MSSY00}, and
an SU(3) symmetry model \cite{MSSY02SU3}, with also three
different options for the contribution from the unfavored
fragmentation functions \cite{MSSY02PLB}. We will show that the
$x$-dependence is sensitive to the ratio of the unfavored
fragmentation function $D_{\bar{u}}^{\Lambda}(z)$ to the favored
fragmentation function $D_{u}^{\Lambda}(z)$, i.e.,
$D_{\bar{u}}^{\Lambda}(z)/D_{u}^{\Lambda}(z)$, whereas the
$z$-dependence is sensitive to the flavor structure of the
fragmentation functions, i.e., the ratio of
$D_u^{\Lambda}(z)/D_s^{\Lambda}(z)$. Future measurements by the
HERMES Collaboration at the Deutsche Elektronen-Synchrotron (DESY)
\cite{SPE} should be able to  discriminate between these various
cases.

We parametrize the quark to $\Lambda$ fragmentation functions
$D^{\Lambda}_q(z)$ by adopting the Gribov-Lipatov relation
\cite{GL}
\begin{equation}
D^{\Lambda}_q(z) \propto q^{\Lambda}(x), \label{GLR}
\end{equation}
in order to connect the fragmentation functions with the quark
distribution functions $q^{\Lambda}(x)$ of the $\Lambda$. More
explicitly, we adopt a general form to relate fragmentation and
distribution functions, as follows \cite{MSSY02PLB}
\begin{equation}
\begin{array}{ll}
D^{\Lambda}_V(z)=C_V(z) z^{\alpha}q^{\Lambda}_V(z),\\
D^{\Lambda}_S(z)=C_S(z) z^{\alpha}q^{\Lambda}_S(z),
\end{array}
\label{dfR}
\end{equation}
where a distinction between the valence ($V$) and the sea ($S$)
quarks is explicit. The above formulae are always correct, since
$C_V(z)$ and $C_S(z)$ are in principle arbitrary functions.  We
should consider Eq.~(\ref{dfR}) as a phenomenological
parametrization for the fragmentation functions of quarks and
antiquarks, as follows
\begin{equation}
\begin{array}{ll}
D^{\Lambda}_q(z)=D^{\Lambda}_V(z)+D^{\Lambda}_S(z),\\
D^{\Lambda}_{\bar{q}}(z)=D^{\Lambda}_S(z).
\end{array}
\label{Dqqbar}
\end{equation}
Three options were found \cite{MSSY02PLB} to fit quite well the
available experimental data of proton production in $e^+e^-$
inelastic annihilation: (1) $C_V=1$ and $C_S=0$ for $\alpha=0$,
(2) $C_V=C_S=1$ for $\alpha=0.5$, and (3) $C_V=1$ and $C_S=3$ for
$\alpha=1$. We adopt these three options to reflect the relation
between unfavored and favored fragmentation functions of the
$\Lambda$.

There is no direct measurement of the quark distributions of the
$\Lambda$. But we can relate the quark distributions between the
proton and the $\Lambda$ by assuming SU(3) symmetry between the
proton and the $\Lambda$ \cite{MSSY02SU3}
\begin{equation}
\begin{array}{lllc}
u_V^{\Lambda}(x)=d_V^{\Lambda}(x)=\frac{1}{6} u_V(x) + \frac{4}{6}
d_V(x),\\
s_V^{\Lambda}(x)=\frac{2}{3} u_V(x) - \frac{1}{3} d_V(x),
\end{array}
\end{equation}
for valence quarks, and
\begin{equation}
\begin{array}{lllc}
\bar{u}^{\Lambda}(x)=\bar{d}^{\Lambda}(x)=\frac{1}{2}
\left[\bar{u}(x)+\bar{s}(x)\right], \\
\bar{s}^{\Lambda}(x)=\bar{d}(x),
\end{array}
\end{equation}
for sea quarks. We adopt the CTEQ parametrization (CTEQ5 set 1)
\cite{CTEQ} of the quark distributions $q(x)$ of the nucleon. In
this way, we get a complete set of quark to $\Lambda$
fragmentation functions, denoted as the SU(3) symmetry model later
on.

It is well known that the flavor structure of $u$ and $d$ quark
distributions of the proton is different between the quark-diquark
model \cite{Fey72,DQM,Ma96} and pQCD based analysis
\cite{Far75,Bro95}: the quark-diquark model predicts that
$d(x)/u(x) \to 0$ at $x \to 1$ whereas a pQCD based approach
predicts that $d(x)/u(x) \to 1/5$. A discrimination between the
two models requires very high precision measurement of the
structure functions at large $x$ and is difficult. On the other
hand, it has been also shown \cite{MSY00PLB} that this flavor
structure of the quark distributions at large $x$ is even more
significant in the case of the $\Lambda$,  with a large difference
between the ratio of $u^{\Lambda}(x)/s^{\Lambda}(x)$: the
quark-diquark model predicts that $u^{\Lambda}(x)/s^{\Lambda}(x)
\to 0$ at $x \to 1$, whereas the pQCD based approach predicts that
$u^{\Lambda}(x)/s^{\Lambda}(x) \to 1/2$. This will produce a large
difference in the ratio of fragmentation functions
$D_u^{\Lambda}(z)/D_{s}^{\Lambda}(z)$, which might be more easily
accessible experimentally via quark to $\Lambda$ fragmentation.

The valence quark distributions of the $\Lambda$ in the
quark-diquark model and the pQCD based analysis have been
explicitly studied \cite{MSY00PLB,MSY00Z,MSSY00} and we adopt the
parametrizations in Ref.~\cite{MSSY00}. To describe the
$\bar{\Lambda}$ fragmentation, it is important to take into
account the sea contributions in the model construction. In order
to use the sea quark distributions from other parametrization
while still keep the flavor structure of the valence quarks as
predicted in the two models, we re-scale the valence quark
distributions by a factor of
$u^{\Lambda}_{V,SU(3)}(x)/u^{\Lambda}_{V,th}(x)$, where the
subscript ``$SU(3)$" denotes the valence quark distributions of
the $\Lambda$ in the SU(3) symmetry model \cite{MSSY02SU3} and
``$th$" denotes the corresponding quantities predicted in the
quark-diquark model or the pQCD based analysis \cite{MSSY00}. This
is done in order to normalize the $\Lambda$ quark distributions to
well known proton quark distribution parametrizations. Notice that
the valence $u$-quark distribution then becomes that of the SU(3)
model, while the others get a rescaling factor. In this way we can
adopt the sea quark distributions from the SU(3) symmetry model as
the sea distributions in the quark-diquark model and the pQCD
based analysis, to reflect the contribution from the unfavored
fragmentation. Thus we get another two sets of quark to $\Lambda$
fragmentation functions, denoted as the quark-diquark model and
the pQCD based analysis later on. This procedure is done with the
main motivation of constructing realistic quark to $\Lambda$
phenomenological fragmentation functions, which have some features
that come from specific theoretical arguments, i.e., the
quark-diquark model and the pQCD based analysis.

\begin{figure}
\begin{center}
\leavevmode {\epsfysize=6cm \epsffile{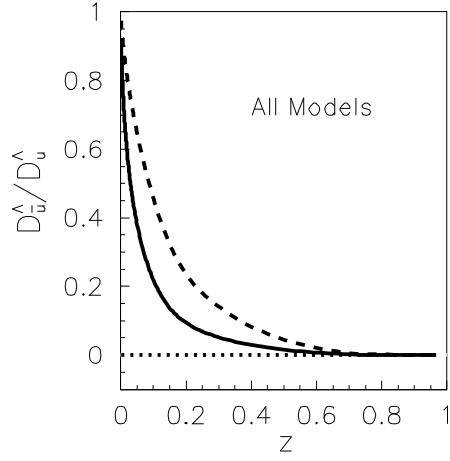}}
\end{center}
\caption[*]{\baselineskip 13pt The ratio
$D^{\Lambda}_{\bar{u}}(z)/D_{u}^{\Lambda}(z)$ of unfavored and
favored fragmentation functions for all three models, with three
options of the contribution from unfavored fragmentation: (1) the
dotted curve with $C_V=1$ and $C_S=0$ for $\alpha=0$, (2) the
solid curve with $C_V=C_S=1$ for $\alpha=0.5$, and (3) the dashed
curve with $C_V=1$ and $C_S=3$ for $\alpha=1$. } \label{msy15f1}
\end{figure}

\begin{figure}
\begin{center}
\leavevmode {\epsfysize=4.8cm \epsffile{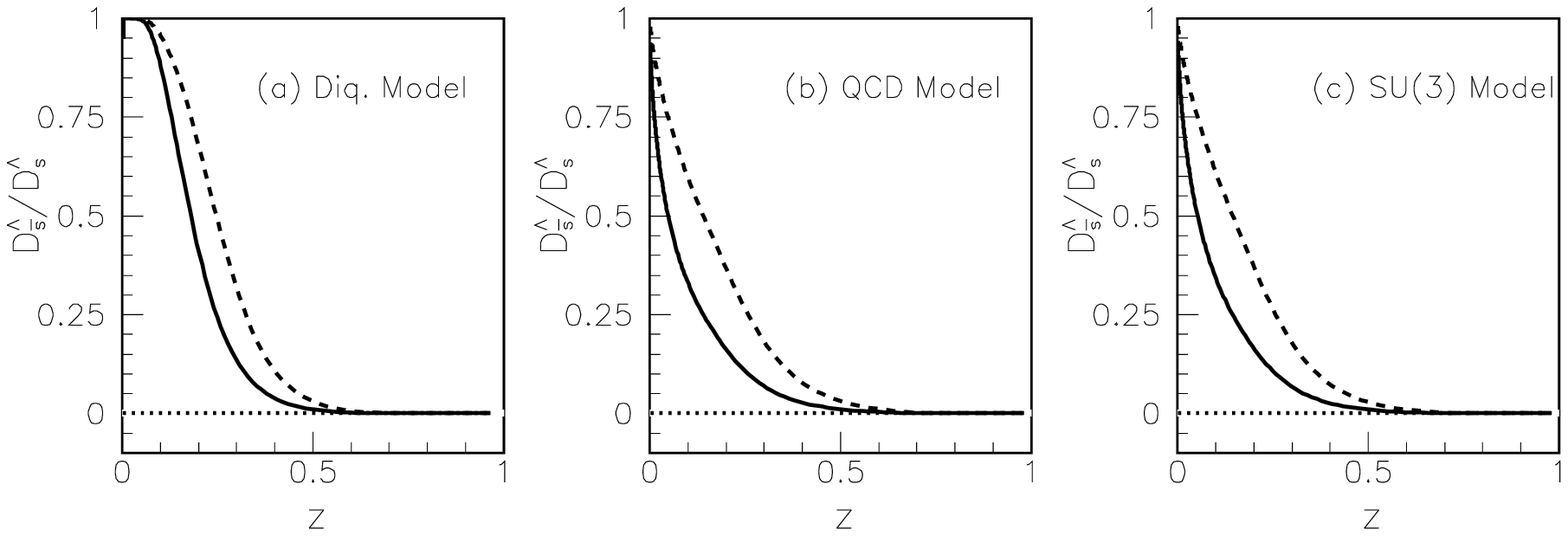}}
\end{center}
\caption[*]{\baselineskip 13pt The ratio
$D^{\Lambda}_{\bar{s}}(z)/D_{s}^{\Lambda}(s)$ of unfavored and
favored fragmentation functions in three different models, with
the dotted, solid, and dashed curves corresponding to the three
options for the unfavored fragmentation as in Fig.~\ref{msy15f1}.}
\label{msy15f2}
\end{figure}

\begin{figure}
\begin{center}
\leavevmode {\epsfysize=4.8cm \epsffile{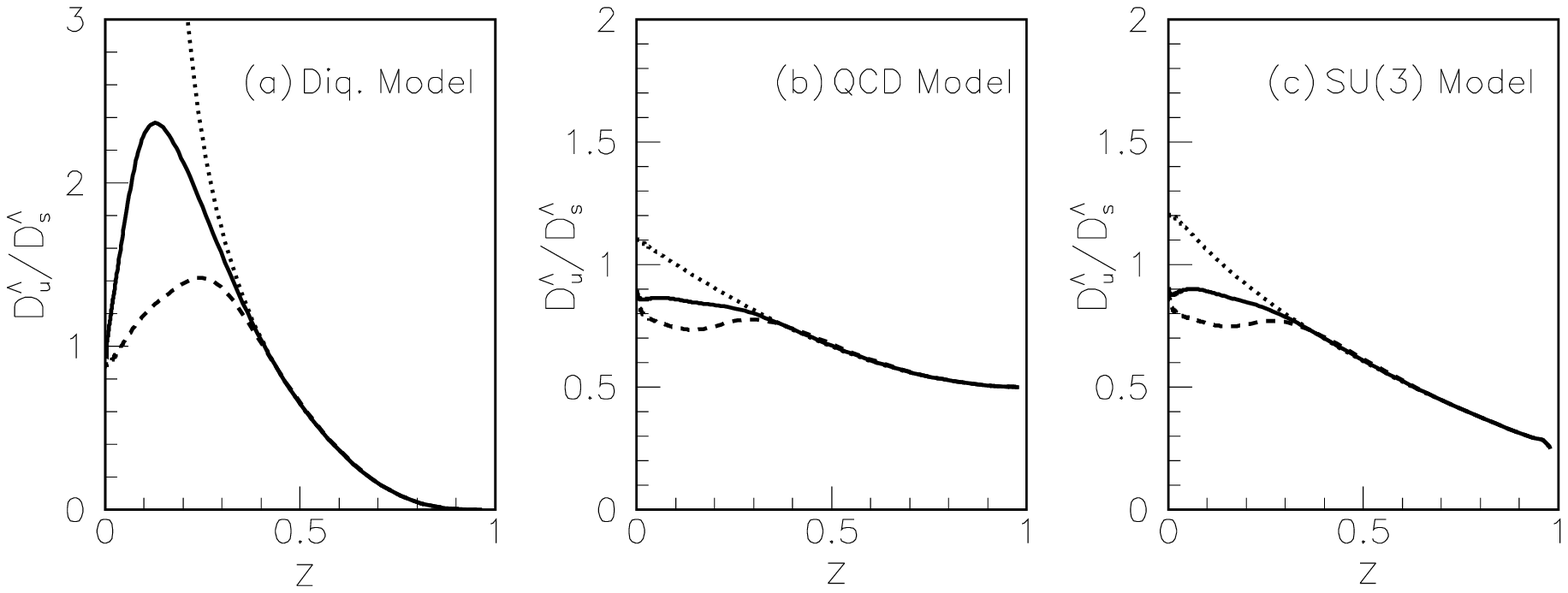}}
\end{center}
\caption[*]{\baselineskip 13pt The ratio
$D^{\Lambda}_{u}(z)/D_{s}^{\Lambda}(z)$ of favored fragmentation
functions in three different models, with the dotted, solid, and
dashed curves corresponding to the three options for the unfavored
fragmentation as in Fig.~\ref{msy15f1}. } \label{msy15f3}
\end{figure}

It can be seen immediately that the ratio of unfavored to favored
fragmentation functions
$D^{\Lambda}_{\bar{u}}(z)/D_{u}^{\Lambda}(z)$ is the same in the
three models:
\begin{equation}
\frac{D^{\Lambda}_{\bar{u}}(z)}{D_{u}^{\Lambda}(s)}=
\frac{D^{\Lambda}_{\bar{u},SU(3)}(z)}{D_{u,SU(3)}^{\Lambda}(z)}.
\end{equation}
We plot this ratio in Fig.~\ref{msy15f1}, and find significant
differences between the three options for the relation between
unfavored and favored fragmentation functions. The contribution
from the unfavored fragmentation is important at small $z$ for
options 2 and 3. We also plot in Fig.~\ref{msy15f2} the ratio
$D^{\Lambda}_{\bar{s}}(z)/D_{s}^{\Lambda}(z)$ in the three models,
and find significant differences between the three options for the
unfavored fragmentation, but with almost similar values in the
three models. In Fig.~\ref{msy15f3}, we plot the ratio of
$D_u^{\Lambda}(z)/D_{s}^{\Lambda}(z)$ in the three models, with
also the three options for the unfavored fragmentation. We find
significant difference of this ratio in the three models:
$D_u^{\Lambda}(z)/D_{s}^{\Lambda}(z) \to 0$ at $x \to 1$ in the
quark-diquark model, whereas $D_u^{\Lambda}(z)/D_{s}^{\Lambda}(z)
\to 1/2$ in the pQCD based analysis and
$D_u^{\Lambda}(z)/D_{s}^{\Lambda}(z) \to 1/4$ in the SU(3)
symmetry model. The flavor structure of the $\Lambda$ differs
significantly at large $z$ in the three models. Both the
quark-diquark model and the pQCD based analysis break the SU(3)
symmetry relation between the proton and the $\Lambda$ with
explicit $z$-dependence.

The differential cross section of $\Lambda$ and $\bar{\Lambda}$
production in semi-inclusive deep inelastic scattering of charged
leptons on a nucleon target can be expressed as
\begin{equation}
\frac{\d^3 \sigma^{\Lambda(\bar{\Lambda})}}{\d x \d y \d
z}=\frac{4\pi \alpha^2 S}{Q^4} \left( 1+(1-y)^2\right)
\sum_{q,\bar{q}} e_q^2 x\, q(x) D_{q}^{\Lambda(\bar{\Lambda})}(z),
\end{equation}
where $x=Q^2/2 M_p \nu$ is the Bjorken scaling variable,
$z=E_{\Lambda(\bar{\Lambda})}/\nu$ is the energy fraction of the
virtual photon energy transferred to the $\Lambda$
($\bar{\Lambda}$), $y=\nu/E_e$ is the fraction of the incident
lepton's energy $E_e$ transferred to the hadronic system by the
virtual photon, $-Q^2=-S x y$ is the squared 4-momentum transfer
of the virtual photon, and $S=M_p^2+m_e^2+2 M_p E_e$ is the
squared 4-momentum sum of the incident lepton and target nucleon
system. For the HERMES experiment, the incident lepton energy
$E_e=27.6~\mbox{GeV}$, and we choose the $x$ range between $0.02
\to 0.5$ with $\left< x\right> =0.09$ for $\Lambda$ and
$\bar{\Lambda}$ events. Therefore we can define the $x$-dependent
ratio of $\bar{\Lambda}/\Lambda$ cross sections
\begin{equation}
R(x)=\frac{\int_0^1\d z\int\d y \frac{\d^3
\sigma^{\bar{\Lambda}}}{\d x \d y \d z}}{\int_0^1\d z\int \d y
\frac{\d^3 \sigma^{\Lambda}}{\d x \d y \d z}},
\end{equation}
and the $z$-dependence ratio
\begin{equation}
R(z)=\frac{\int_{0.02}^{0.5}\d x\int\d y \frac{\d^3
\sigma^{\bar{\Lambda}}}{\d x \d y \d z}}{\int_{0.02}^{0.5}\d x\int
\d y \frac{\d^3 \sigma^{\Lambda}}{\d x \d y \d z}}.
\end{equation}
We calculate these ratios by adopting the CTEQ parametrization of
the quark distributions for the proton target and the above three
model results for the quark to $\Lambda$ fragmentation functions.

\begin{figure}
\begin{center}
\leavevmode {\epsfysize=4.8cm \epsffile{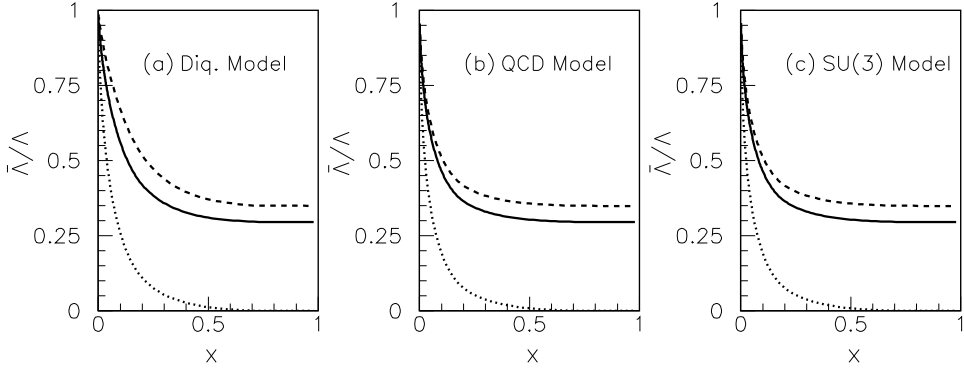}}
\end{center}
\caption[*]{\baselineskip 13pt The $x$-dependence of the
$\bar{\Lambda}/\Lambda$ ratio $R(x)$ in three different models,
with the dotted, solid, and dashed curves corresponding to the
three options of the unfavored fragmentation as in
Fig.~\ref{msy15f1}. } \label{msy15f4}
\end{figure}

The $x$-dependence of the  $\bar{\Lambda}/\Lambda$ ratio $R(x)$ is
plotted in Fig.~\ref{msy15f4}, from which we find significant
differences between the three different options of the unfavored
fragmentation, but almost no obvious differences between the three
models. This can be understood intuitively, as the $\Lambda$ and
$\bar{\Lambda}$ events are dominated by small $z$ contribution in
the integration over $z$, and the ratio $R(x)$ is predominantly
determined by the ratio
$D_{\bar{u}}^{\Lambda}(z)/D_{u}^{\Lambda}(z)$ at small $z$. Thus
the $x$-dependence of the $\bar{\Lambda}/\Lambda$ ratio is
sensitive to the ratio of the unfavored fragmentation function
$D_{\bar{u}}^{\Lambda}(z)$ over the favored fragmentation function
$D_{u}^{\Lambda}(z)$. Therefore $R(x)$ is a sensitive physical
quantity to provide a discrimination between different options of
the contribution from the unfavored fragmentation.

\begin{figure}
\begin{center}
\leavevmode {\epsfysize=4.8cm \epsffile{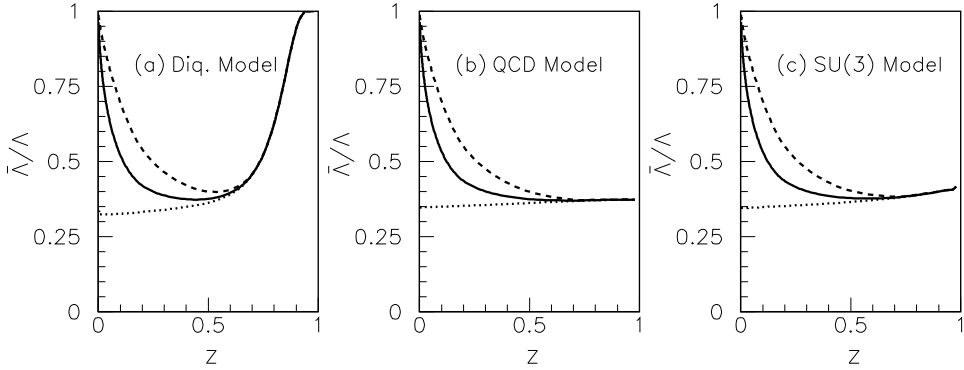}}
\end{center}
\caption[*]{\baselineskip 13pt The $z$-dependence of the
$\bar{\Lambda}/\Lambda$ ratio $R(z)$ in three different models,
with the dotted, solid, and dashed curves corresponding to the
three options of the unfavored fragmentation as in
Fig.~\ref{msy15f1}. } \label{msy15f5}
\end{figure}

\begin{figure}
\begin{center}
\leavevmode {\epsfysize=4.8cm \epsffile{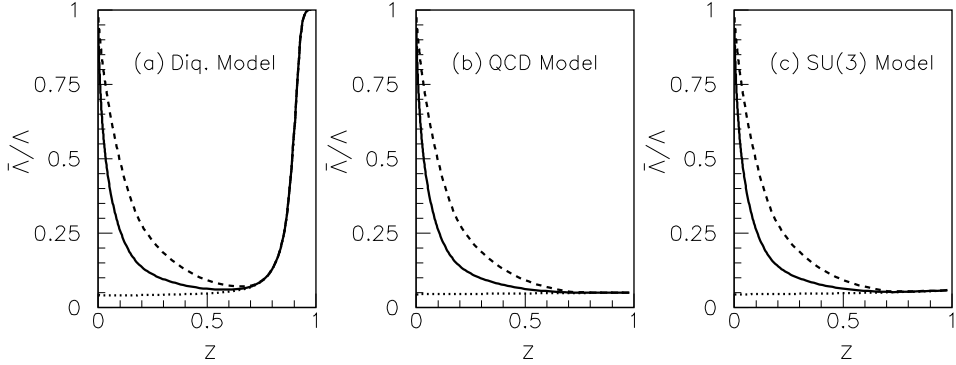}}
\end{center}
\caption[*]{\baselineskip 13pt Same as Fig.~\ref{msy15f5}, but the
integrated $x$ range is $\left[0.2,0.5\right]$, rather than
$\left[0.02,0.5\right]$ as in  Fig.~\ref{msy15f5}.  }
\label{msy15f6}
\end{figure}

The $z$-dependence of the $\bar{\Lambda}/\Lambda$ ratio $R(z)$ is
plotted in Fig.~\ref{msy15f5}, from which we find significant
difference between the quark-diquark model and the other two
models. We notice that the ratio $R(z)$ increases rapidly as $z$
increases in the quark-diquark model. This can be explained by the
fact that the $s$-quark to $\Lambda$ ($\bar{s}$-quark to
$\bar{\Lambda}$) fragmentation dominates over the fragmentation of
$u$ ($\bar{u}$) and $d$ ($\bar{d}$) quarks at large $z$. The small
$x$ region $\Lambda$ ($\bar{\Lambda}$) events dominate in the
integration over $x$, where the relatively small number of $s$
($\bar{s}$) quarks inside the nucleon target cannot be neglected.
Thus the ratio $R(z)$ at $z\to 1$ is predominantly determined by
the ratio $\bar{s}/s$ of the proton at small $x$. However, for the
pQCD based analysis and the SU(3) symmetry model, the
fragmentation to $\Lambda$ ($\bar{\Lambda}$) from $u$ ($\bar{u}$)
and $d$ ($\bar{d}$) quarks are only reduced by a factor of 2 and 4
respectively in relative to the $s$-quark to $\Lambda$
($\bar{s}$-quark to $\bar{\Lambda}$) fragmentation at large $z$,
thus the ratio of $R(z)$ at $z\to 1$ is predominantly determined
by the ratio $\bar{u}/u$ of the proton at small $x$. The explicit
value of the ratio is sensitive to the $x$-integrated range, but
the qualitative features of the different models remain unchanged,
as shown in Fig.~\ref{msy15f6}. Therefore $R(z)$ is a sensitive
physical quantity that can discriminate between different flavor
structures reflected by the ratio
$D_u^{\Lambda}(z)/D_{s}^{\Lambda}(z)$ of favored fragmentation
functions.

Although the Gribov-Lipatov realtion that we have used in this
work is strictly valid only in the limit $ z\to 1$ \cite{GL},  it
provides a reasonable effective guidance for a phenomenological
parametrization of the quark to $\Lambda$ fragmentation functions,
and in this sense it is a useful but not essential ingredient of
our analysis. Of course, the fact that $R(z)$ gives information
about the flavor-dependent fragmentation functions
$D_{s}^{\Lambda}(z)$ and $D_u^{\Lambda}(z)$ is independent on the
Gribov-Lipatov relation.

In summary, we find in this letter that the
$\bar{\Lambda}/\Lambda$ cross section ratio of semi-inclusive
$\Lambda$ and $\bar{\Lambda}$ productions in deep inelastic
scattering of charged leptons on a nucleon target is a sensitive
physical quantity that can provide useful information on the quark
to $\Lambda$ fragmentation functions. The $x$-dependence of the
ratio can discriminate between different options of the
contribution from the unfavored fragmentation, whereas the
$z$-dependence of the ratio can provide a sensitive discrimination
between different flavor structure of the favored
$D_u^{\Lambda}(z)$ and $D_{s}^{\Lambda}(z)$ fragmentation
functions. Thus measurements of the ratio $\bar{\Lambda}/\Lambda$
by the HERMES Collaboration will provide important information on
the relations between different flavor-dependent quark to
$\Lambda$ fragmentation functions.

{\bf Acknowledgments }

This work is partially supported by National Natural Science
Foundation of China under Grant Numbers 10025523 and 90103007, by
Fondecyt (Chile) grant 1030355, by Alexander von Humboldt-Stiftung
(J.-J.~Yang), and by Foundation for University Key Teacher by the
Ministry of Education (China).

\end{document}